\documentclass[aritcle,aps,epsf,amsfonts,amssymb,amsmath,nofootinbib]{revtex4}

\usepackage{amsmath}
\usepackage{latexsym}
\usepackage{empheq}
\usepackage{amssymb}
\usepackage{titlesec}
\usepackage[titletoc,toc,title]{appendix}
\usepackage{graphics}
    \usepackage{appendix}

\usepackage{amsfonts}

\titleformat*{\subsubsection}{\large\itshape}

\begin{document}

\begin{center}

\thispagestyle{empty}
\pagestyle{empty}

{ \huge \bfseries Bessel Functions in Mass Action} \\
[0.5cm]

{\huge \bfseries Modeling of Memories and Remembrances }\\[1.5cm]

\begin{minipage}{1\textwidth}
\begin{flushleft} \large

\large Walter J. \textsc{Freeman${}^{1}$}, Antonio \textsc{Capolupo${}^{2}$ }, Robert \textsc{Kozma${}^{3}$ }
 Andr\'es \textsc{Olivares del Campo${}^{4}$}, Giuseppe \textsc{Vitiello${}^{2}$\footnote{corresponding author; vitiello@sa.infn.it}}.

\vspace{1.0cm}

${}^{1}$ Department of Molecular and Cell Biology,
University of California, Berkeley CA 94720-3206
USA\\

${}^{2}$ Dipartimento di Fisica E.R.Caianiello
 Universit\'a di Salerno, and INFN Gruppo collegato di Salerno, Fisciano (SA) - 84084, Italy

${}^{3}$ Department of Mathematics, Memphis University, Memphis TN 38152 USA

${}^{4}$ The Blackett Laboratory, Imperial College London, Prince Consort Road, London SW7 2BZ, UK
\vspace{1.5cm}
\end{flushleft}
\end{minipage}

{\large\bfseries Abstract}\\
\end{center}
\small

Data from experimental observations of certain brain functions, such as impulse responses of cortex to electric shocks (average evoked potentials), related with the coexistence of pulses and wave modes in the dynamics of a class of neurological processes (Freeman K-sets), present functional distribution reproducing the Bessel function behavior.
This suggests the possibility to replace ordinary differential equations, typically used to model
data bases concerning such processes, with couples of damped/amplified oscillators which provide time dependent representation of spherical Bessel equation.
The root loci of poles and zeros of the equation solutions are shown to conform to solutions of K-sets. One advantage of the present formalism is that some light is shed on the problem of filling the gap between the behavior at cellular level and the macroscopic dynamics involved in the traffic between the brain and its environment. Breakdown of time-reversal symmetry in each of the (damped and amplified) oscillators is related with the cortex thermodynamic features. This is proposed to be a possible mechanism to deduce lifetime of recorded memory.

\vspace{0.5cm}

\section{Introduction}

There is an essential problem in the study of brain function that even today, after so many years since Karl Lashley posed his dilemma, still waits for a solution. As recalled many times in the literature, in the mid 1940s he wrote~\cite{Lashley}:

" .... Here is the dilemma. Nerve impulses are transmitted ...from cell to cell through definite intercellular connections. Yet, all behavior seems to be determined by masses of excitation...within general fields of activity, without regard to particular nerve cells... What sort of nervous organization might be capable of responding to a pattern of excitation without limited specialized path of conduction? The problem is almost universal in the activity of the nervous system."

The successes of neuroscience in the study of the structural and biochemical properties of neurons, glia cells, and all the biological units and cellular structures in the brain have not yet filled the gap between the behavior quite well understood at cellular level and the macroscopic dynamics involved in the traffic between the brain and the world around it. Much work is still to be done in order to have better and better insight in a matter where the connection between phenomenological data and a reliable theoretical representation is still facing extreme difficulties. Within the limits discussed below, this is the problem on which we focus our attention in this paper.

In order to analyze it,  we consider data from experimental observations made by fitting measurements of brain function with solutions to ordinary differential equations (ODE) (Freeman K-sets)~\cite{Freeman1975}. From the observations it emerges that data from neurological processes, such as impulse responses of cortex to electric shocks (average evoked potentials (AEP)), present functional distribution in plots of amplitude (in $\mu Volts$) {\it vs} time (in {\it msec}) as shown in
(Fig.~1)~\cite{Freeman1975} reproducible by the Bessel function behavior.
Since Bessel equations have a time dependent representation in terms of couples of equations, each couple containing one equation for a damped oscillator, the other one for an amplified oscillator, both with time dependent frequency, we are motivated to study the observed neural processes in terms of these oscillators, which is also consistent with the fact that consideration of dynamics in the continuum of the wave mode cannot be omitted, as it appears from the experimental observations we start with, as well as from other data and
analysis~\cite{FreemanQuiran,Robinson,Bressloff} (see also Appendix A) showing that pulses and wave modes coexist simultaneously in the dynamics of neural populations, whose understanding thus requires to consider both of them.

The behavior of each parametric oscillator emerges from an interactive population of excitatory cells coupled with an interactive population of inhibitory cells forming a negative feedback loop, in which the two populations are in quadrature having the same frequency and decay rate, with on average the excitatory oscillation leading the inhibitory oscillation by $\pi/2 ~radians$ (cf. Appendix A.3)\footnote{The phase lead may be varied from average by the two kinds of positive feedback in the populations if they are not equal to the negative feedback gain.}. The damped oscillator has low feedback gain; the amplified oscillator has high feedback gain. We obtain, as a result of our study, the agreement between the dissipative dynamics expressed by each oscillator in the couples and the conclusions on the root loci and the evolution towards limit cycles presented in~\cite{Freeman1975} (see Section IV and Appendix B). Our analysis in terms of the parametric oscillator couples is also consistent with temporal frequency modulation as it is derived from observations~\cite{Freeman1975,FreemanQuiran} and widely used in the literature~\cite{Robinson,Bressloff, Plenz,Puljic} .
In addition to this, we also obtain a link with
the dissipative many-body model of brain~\cite{MyDouble,PLR},  since the couples of damped/amplified oscillators is known to have a representation in terms of correlation modes in many-body systems~\cite{MyDouble,PLR,Vitiello1995,CurrentNeuro}. Each couple forms an operator and its double, one running forward in time in the brain, the other backward in time~\cite{MyDouble} in the mind.

Our analysis is limited to the data mentioned above which are well fitted by the Bessel function. Perhaps, a plausible macroscopic physical justification of the observed similarity of evoked potentials with Bessel functions is likely due to the 2-dimesionality of the distribution of the excitatory and inhibitory neurons. On the other hand, the occurrence of Bessel functions might be not so surprising since in physics Bessel functions play a crucial role in theoretical descriptions of (resonant) pulses in cross sections of particle scattering as well as of wave propagation processes, which also suggests that the observed brain functions have, indeed, a lot to do, simultaneously, with pulse modes and wave modes, as already mentioned.
The representation in terms of couples of damped/amplified oscillators also allows the interplay of linearity/nonlinearity dynamics features~\cite{Andronov}. Such an interplay is of crucial importance in brain studies (see our comments in closing Section II). Linearity allows us to consider the summations in the fitting functions, as we will see in the following Sections, and it has been one of the motivations to adopt the analysis in terms of ODE in \cite{Freeman1975}. On the other hand, the  biophysical nonlinearity at the microscopic level~\cite{PLR} appears at the macroscopic level as the root loci~\cite{Freeman1975} revealing the capability by neural activity to provide the transition energy necessary to sustain the phase transition. Such a nonlinearity is present in our modeling since the  damped/amplified oscillators are actually nonlinearly coupled as it appears considering their dynamical properties~\cite{Andronov,Pashaev}.
We will comment more on this point in Section II.

In this paper, we do not consider plots of evoked potential data showing functional distributions different from Bessel functions. However, since any function can be expanded in a series of Bessel functions (see e.g.~\cite{Abramowitz,Messiah,Arfken}), the representation in terms of damped/amplified oscillators and our results can be extended also to such cases, within quite general boundary conditions and provided a convenient analysis is carried on, which we plan to do in a future work.

The functional activity of the brain and the laboratory observations at the cellular (neuronal) level represent the manifestation of the dynamics underlying at the elementary constituent level. The degrees of freedom of such a dynamics are associated with the elementary components and modes of the molecules present in the system, such as the electrical dipole vibrational modes of molecules of biological interest and of the water molecules constituting the bath in which they are embedded. These are of course quantum degrees of freedom. Their dynamics is studied in the dissipative quantum model of brain~\cite{Vitiello1995}, which deals indeed with the quantization of the damped/amplified couples of oscillators. We do not consider the dissipative model in this paper. In any case, we stress that neurons and glia cells are considered to be classical objects in this paper and in the dissipative model.

The plan of the paper is the following. In Section II we outline the derivation of the time dependent representation of the Bessel equation in terms of the damped/amplified couples of oscillators.
We present in the Appendices A the details of the empirical setting in which the data analyzed in this paper were derived.
Section III is devoted to the analysis of the data in terms of the couple of damped/amplified oscillator equations.
Applications and explicit examples of computation in our formalism are discussed in Section IV, with formal details given in the Appendix B. In Section V, a hierarchy of lifetimes for k-mode (k is a wave number) components is considered. Section VI is devoted to conclusions.
For the sake of brevity, in this paper, we do not discuss features such as scale free behaviors in brain functioning, which may be analyzed in in terms of coherent state dynamics in the dissipative model. For a discussion on this last point see Refs.~\cite{NMNC,PLA2012}.

\begin{figure}[hbtp]\label{fig:Freeman1975}
\centering
\resizebox{11cm}{!}{\includegraphics{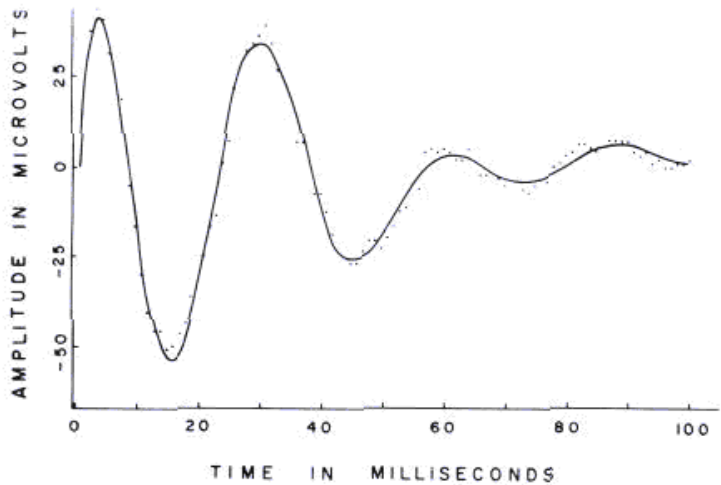}}
\caption{\small \noindent AEP and fitted curve for impulse response of the prepyriform cortex in the closed loop state. Picture adapted from Ref.~\cite{Freeman1975}, Section II, Page 111.}
\end{figure}

\section{Bessel equations and the damped/amplified oscillator couples}

A conclusion that can be drawn from the laboratory data and their analysis, as summarized in the Appendix A where the empirical setting and the scenario in which the data were collected are presented, is that the dynamics of neural populations cannot be understood from either waves or pulses alone. As well known in the literature~\cite{Freeman1975,FreemanQuiran,Schomer}, pulses and wave modes indeed coexist simultaneously in circular causality. Each determines the other. Laboratory observations are made by batch processing, with categorization after each shot. What has been  found~\cite{Freeman1975} is that, in the successive brief time windows accessible to observations, the dynamics of dendritic integration and axonal transmission by neural populations combine the two modes. On the one hand, sometimes the paradigm prevails which uses only data from the pulse mode. It omits consideration of dynamics in the continuum of the wave mode and focuses on the point processes revealed by action potentials. On the other hand, one cannot omits consideration of dynamics in the continuum of the wave mode.
The above-mentioned laboratory observation that functions representing neurobiological processes can be  fitted by Bessel functions is significant for opening to investigation the properties of neural collectives.
We consider thus the spherical Bessel equation of order $n$~\cite{Abramowitz,Messiah,Arfken}:
\begin{equation}
z^2\frac{d^2}{dz^2}M_n + 2z\frac{d}{dz}M_n + [z^2-n(n+1)]M_n=0 ~, \label{(2)} \\
\end{equation}
with $n$ being zero or an integer. Equation (\ref{(2)}) admits as particular solutions Bessel functions of the first and second kind. Superposition of both is called a Hankel function~\cite{Abramowitz,Messiah,Arfken}.

Note that a transformation of the type $n \rightarrow -(n+1)$ leaves Equation (\ref{(2)})  invariant so that both $M_n$ and $M_{-(n+1)}$ are solutions of the same equation. This degeneracy can be lifted by introducing the parametrization as done in~\cite{AlfinitoVitiello2000}:
$q_{n,l}=M_n \cdot (e^{-\frac{t}{\alpha_n}})^{-l}$, $z \equiv z (t)=\epsilon_n \, e^{-\frac{t}{\alpha_n}}$,
where $\epsilon_n$ and $\alpha_n$  are arbitrary parameters and $t$ denotes the time variable. $\epsilon_n$ is dimesionless and $\alpha_n$ has the dimension of time.
By setting $n(n+1)=l(l+1)$,  one obtains two differential equations for $l=-(n+1)$ and $l=n$, respectively, where the symmetry under the transformation $n \rightarrow -(n+1)$ is broken~\cite{AlfinitoVitiello2000}. These differential equations can be reduced to the parametric damped/amplified oscillator equations:
\begin{eqnarray}
 \ddot{\xi} + L\dot{\xi} + \omega_{n}^2(t)\,\xi &=& 0 ~,  \label{(3a)}  \\
 \ddot{\zeta} -L \dot{\zeta} + \omega_{n}^2(t)\, \zeta &=& 0  ~,  \label{(3b)}
\end{eqnarray}
where the dots represent derivatives with respect to time,  L is a parameter intrinsic to the system (e.g. a biological, physiological or behavioral parameter) and the shorthand notation is: $\xi \equiv  q_{n,-(n+1)}$\,, $\zeta \equiv  q_{n,n}$\,, $L  \equiv  \frac{2n+1}{\alpha_{n}} $. The time-dependent frequencies arising after introducing the above parametrization is $\omega_{n}(t)= \frac{\epsilon_n}{\alpha_n}e^{-\frac{Lt}{2n+1}}=\omega_0\,e^{-\frac{t}{\alpha_n}}$. Note that $\omega_{0}\equiv \frac{\epsilon_n}{\alpha_n}=k c \equiv \omega_{0,k}$, with $k$ the wave-number $k = 2\,\pi/\lambda$. The $k$ dependence of $L$ can be expressed as $L_k=\frac{(2n+1)}{\epsilon_n} k c$. This is done in such a way that $\epsilon_n$, $L_k$ and ${\omega}_{0,k}$, which have dimension of $t^{-1}$, are independent of $n$ (fixed time-invariant frequencies).
Note that the transformation $n \rightarrow - (n+ 1)$ leads
to solutions (corresponding to $M_{- (n+1)}$) which can be obtained  by time-reversal $t \rightarrow - t$ in
$\omega_{n}(t)$. We will omit the $k$ index for simplicity whenever no misunderstanding arises.

Eqs.~(\ref{(3a)}) and (\ref{(3b)})  are recognized to be nothing else than the couple of equations at the basis of the
dissipative model of brain~\cite{Vitiello1995} with time-dependent frequency $\omega (t)$.
They can be reduced to the single parametric oscillator
\begin{flalign}
 &\ddot{r}_n  + \Omega_{n}^2(t)r_n = 0 ~,  \label{(4)}
\end{flalign}
where $q_{n,-(n+1)}=\frac{1}{\sqrt{2}}r_ne^{-\frac{Lt}{2}},q_{n,n}=\frac{1}{\sqrt{2}}r_ne^{\frac{Lt}{2}}$ and $\Omega_n$ is the common frequency of oscillation:
\begin{flalign}
 &\Omega_{n}(t)=\Big[\omega_n^2(t)-\frac{L^2}{4}\Big]^\frac{1}{2}  \ge 0 ~, \label{(5)}
\end{flalign}
so that the functions $q_{n,n}$ and $q_{n,-(n+1)}$
are ``harmonically conjugated''. Proof of the mechanism is that the signals are observed to be in quadrature; the frequency and decay rate are always identical, and the phase of the inhibitory signal lags a quarter cycle behind the excitatory signal. Note that $\Omega(t)$ becomes time-independent as $t\rightarrow \infty$ and it is fully specified by characteristic parameters of the system\footnote{Gamma bursts of oscillation in the ECoG lasting long enough ($ >3$ to $5$ cycles) commonly show temporal frequency modulation (smooth decrease of $10\,\%$ of the mean, but rarely an increase). The impulse responses (average evoked potentials) has fixed time-invariant frequencies. This is equivalent to time $t$ going to infinity.}.
$\Omega$ is assumed to be real in order to avoid the overdamping regime.

In conclusion, the Bessel equation (\ref{(2)}), for each $n$, has a time dependent representation in terms of  the  Eqs.~(\ref{(3a)}) and (\ref{(3b)}) of the couple of damped/amplified oscillators, provided the above parametrization is adopted.

We observe that in the dissipative quantum model of brain Eqs.~(\ref{(3a)}) and (\ref{(3b)}) describe electric dipole vibrational modes, denoted by $a_k$ and $\tilde{a}_{k}$, respectively,   called dipole wave quanta (DWQ), with the suffix $k$ denoting the wave-number introduced above. These are the Nambu-Godstone (NG) boson quanta dynamically generated as a result of the breakdown of the rotational symmetry of the molecule electrical dipoles~\cite{MyDouble,PLR,Vitiello1995,CurrentNeuro,FreemanVitiello2009}. Their condensation in the ground state of the system leads to the formation of ordered (coherent)  patterns~\cite{Celeghini1992,Alfinito1999,AlfinitoManka2000}. The brain couples to its environment (referred to as its \emph{Double}~\cite{MyDouble,PLR,Vitiello1995,CurrentNeuro,FreemanVitiello2009})  through the perception of external stimuli. These produce the symmetry breakdown and trigger the condensation process. This is constrained by the requirement $\cal{N}$$_{a_k}$$-\,$$\cal{N}$$_{\tilde{a}_{k}}=0$, with  ${\cal{N}}_{a_k}$ and ${\cal{N}}_{\tilde{a}_{k}}$ denoting the number of ${a_k}$ and ${\tilde{a}_{k}}$ modes, respectively, so that the  ground state has the lowest energy. Such a constrain does not fix, however, the number of $k$-modes, for each $k$, in the ground state and we can have infinitely many lowest energy states with distinct combinations of different momentum modes, characterized by the ``order parameter'' ${\cal{N}}   \equiv \{{\cal{N}}_{{a}_{k}} = {\cal{N}}_{\tilde{a}_{k}}, \,\forall \, k \}$. The ${\cal{N}}$-set  thus labels the system ground states (referred to as the the memory states in the dissipative model).

We close this Section with two comments. The first one is about the breakdown of time-reversal symmetry introduced by Eqs.~(\ref{(6)}) and  (\ref{(8)}). This is intrinsic to the dissipative/amplified character of Eq.~(\ref{(3a)}) and  Eq.~(\ref{(3b)}). Breakdown of time-reversal symmetry introduces the  ``arrow of time'' (in each of the damped and amplified oscillators), which, as discussed in Refs.~\cite{AlfinitoVitiello2000,FreemanVitiello2009,Celeghini1992,Alfinito1999,AlfinitoManka2000}, can be related to the ``thermodynamical'' and the ``psychological'' arrows of time. Among the many existing papers dealing with nonequilibrium thermodynamics and  time-reversal symmetry breakdown in neuronal systems we only quote here few of them, see~\cite{Taniguchi,Badel,Paninski,Wilson,Ingber}. We do not insist further on this topic in this paper.

The other comment concerns the fact that the system of equations~(\ref{(3a)}) and (\ref{(3b)}) allows to us to consider the summations in the fitting functions, as we will see in the following Section, cf.  Eqs.~(\ref{(6)}) and  (\ref{(8)}).
Linearity,  which, as recalled in the Introduction, has been one of the motivations to adopt the analysis in terms of ODE in \cite{Freeman1975},  is thus implicit also in the present treatment. For observational analysis is highly convenient that the generic equation for the impulse response in time and space is the sum of the responses in the two modes of observation (pulse and wave mode). This motivates the choice of the fitting functions  used in Ref.~\cite{Freeman1975} and  considered in the next Section, see Eq.~(\ref{(6)}) (and  (\ref{(8)})). On the other hand, biophysical nonlinearity  appearing at the macroscopic level as the root loci~\cite{Freeman1975} reveals the capability by neural activity to provide energy necessary to sustain phase transition resembling a Hopf bifurcation. It is well known that nonlinearity is crucial in brain, especially in the pulse-wave transit asymmetric sigmoid function~\cite{Ilin1,Ilin2}. Such a nonlinearity is also present in our modeling since the  damped/amplified oscillators are known to be nonlinearly coupled~\cite{Andronov}, $L$ being the dissipation/loading parameter in the combined damped/amplified system dynamics, as it appears also from the entangled two modes quantum coherent state they generate at quantum level
(the interested reader is referred to \cite{Vitiello1995,FreemanVitiello2009} for details). Moreover, the root loci studied in Ref.~\cite{Freeman1975}, and in the literature therein quoted, also appear in the present treatment, as we will show in the following Sections. It is also interesting that the time independent value $\omega_0$ of the frequency $\omega_{n}(t)$ is reached for higher values of $n$,  namely in the limit of the maximal coupling between the
damped and the amplified  oscillators (the whole system $(\xi, \, \zeta)$
is a closed system in that limit of large $n$. Considering that a brain is a thermodynamic system with a fuzzy number of ports, n as an integer must be very large indeed.).  This is the limit of stationarity: observations show that most cortical impulse responses (evoked potentials from single shock electrical stimulation) lasted only $1.5$ to $2.5$ cycles, and those few lasting longer had fixed stationary frequency. The linear range is crucial for the test of superposition (which allows the summation in the fitting functions  (\ref{(6)}) and  (\ref{(8)})) since there the pulse density can be represented as a function of wave density through the $G(p)$ and/or $G(v)$ functions (cf. Eqs.~(\ref{2}) and (\ref{3}) and Ref.~\cite{Freeman1975}). In order to use time ensemble averaging to estimate pulse density in the linear range, one also assumes ergodicity, which can be done also in the present case. Thus, linearity, stationarity and ergodicity, combined with nonlinearity, are compatible with the damped/amplified parametric oscillator treatment.

Our next task is to show how the formalism concretely works in explicit examples.

\section{Fitting functions and the Bessel equation}

In this section we show that neuronal processes described in Ref.~\cite{Freeman1975} can be studied by using the damped/amplified oscillator equations along the lines discussed above leading to results in agreement with those obtained in~\cite{Freeman1975}. In Section IV and V we then present explicit examples of computations (see also the Appendix B) and discuss the lifetime of  correlation modes ($k$-modes), respectively, in order to show how the present scheme concretely works.

The fitting functions for the laboratory observations used in \cite{Freeman1975} (see in particular Section 2.5.3)  to describe the oscillatory processes resulting from connections between sets of neurons  are generally of the form
\begin{flalign}
&v(t)=\sum_{j}^n A_j\,e^{-p_j\,t} ~, \label{(6)}
\end{flalign}
where $A_j,\, p_j \in \mathbb{C}$ with positive coefficients (i.e. $Re(p_j) \ge 0 $ and $Im(p_j) \ge 0$). When $p_j$ is real, $A_j$ is also real. Otherwise, $p_j$ and $A_j$ always come in conjugate pairs to yield damping $\sin$ or $\cos$ terms of the type $\sin{\big(Im(p_j)\,t + \varphi_j\big)}e^{-Re(p_j)\,t}$ with $\varphi_j$ being a constant phase term in $A_j$. For example, using the Euler identity $2\sin \theta = i\,e^{-i\,\theta}- ie^{i\,\theta}$, with `angle' $\theta$, the term $\sin(\Gamma_j\, t + \varphi_j)\,e^{-\gamma_j\,t}$ can be rewritten as $A_je^{-p_j\, t} + A_j^*\,e^{-p_j^*\,t}$ where $A_{j}=e^{-i(\varphi_{j}-\frac{\pi}{2})}/{2}$, $p_{j}=\gamma_{j} + i \Gamma_{j}$ and similarly for a damping $\cos$ term. We observe that  $A_j$ and $p_j$ actually depend on the wave-number $k_j$.  In the field theory frame adopted by us,   $k_j$, for each $j$, varies over a continuum set of values. We will use the simplified notation $A_j \equiv A_{k_j} $ and $p_j \equiv p_{k_j}$ whenever no misunderstanding arises.  The Laplace transform of Eq.~(\ref{(6)}) is
\begin{flalign}
&V(s)= \frac{\prod_{i}^m(s+ z_i)}{\prod_{j}^n(s+p_j)} ~,\label{(7)}
\end{flalign}
which has the zeros at the at $z_i$ and all its poles lying in the left-hand side of the complex $s$-plane.

Each term in the summation in Eq.~(\ref{(6)}) corresponds to a DWQ $a_{k_j}$ mode of the dissipative quantum model of brain. This is suggested by the fact that, as we will explicitly show, each term in Eq.~(\ref{(6)}) satisfies Eq.~(\ref{(3a)}). Hence, the modeled output potential is described by a linear superposition of $a_{k_j}$ modes which represent solutions of the $n$-order Bessel equation. In this way, a particular stimulus will excite an $\cal{N}$-set which is made up of different $\cal{N}$$_{a_k}$ components~\cite{FreemanVitiello2009}.

Moreover, one should remark that, for each $j$, the complete basis of functions, of which $e^{-p_j\, t}$ is one element, is  $\{e^{-p_j\, t}, \, e^{+p_j t} \}$. This means that one cannot avoid to consider also the element $e^{+p_j t}$. In other words, mathematical completeness  requires that we need to ``double'' the degrees of freedom of the system by introducing  the other functions of the basis, $e^{+p_j\, t}$, and thus we also need to consider
\begin{flalign}
&\tilde{v}(t)=\sum_{j}^n A_j\,e^{p_j \,t} ~,\label{(8)}
\end{flalign}
where again $A_j,\, p_j$ are complex numbers with positive coefficients. These terms are solutions of Eq.~(\ref{(3b)}) associated to the $\tilde{a}_k$ modes. This shows the consistency between the mathematical necessity of considering the complete functional basis $\{e^{-p_j\, t}, \, e^{+p_j t} \}$ and the derivation in Section II of the couple of damped and amplified oscillator equations. It is worth noting that the Laplace transform of Eq.~(\ref{(8)}) is
\begin{flalign}
&\tilde{V}(s)= \frac{\prod_{i}^m(s+ z_i)}{\prod_{j}^n(s-p_j)} ~, \label{(9)}
\end{flalign}
which has poles lying in the right hand side of the complex $s$-space.
Restoring the $k$ subscript notation, use of $A_{j} (t) = A_{j}\,e^{-{p_j}\, t}$, $A_{j}\neq 0$,  in  Eq.~(\ref{(3a)}) gives:
\begin{flalign}
 &p_{k_j}^2 - L_{k_j}p_{k_j} + \omega_{n,k_j}^2(t) = 0 ~, \label{(10)}
\end{flalign}
so that
\begin{flalign}
p_{k_j} = \frac{ L_{k_j}}{2} \pm i \Omega_{k_j,n}(t) ~, \label{(11)}
\end{flalign}
for real coefficients.
For complex coefficients we obtain:
\begin{flalign}
 &\gamma_{k_j}^2 - \Gamma_{k_j} ^2 + 2i(\gamma_{k_j} \Gamma_{k_j}) - L_{k_j}(\gamma_{k_j} + i\Gamma_{k_j}) + \omega_{n,k_j}^2(t)= 0~. \label{(12)}
\end{flalign}
After separating real and imaginary parts,
\begin{eqnarray}
  L_{k_j} &=& 2\gamma_{k_j}\label{(13)} ~, \\
 \omega_{n,k_j}^2(t)  &=& \Gamma^2_{k_1} + \gamma^2_{k_1} = |p_{k_j}|^2 ~,\label{(14)}~
\end{eqnarray}
and rearranging:
\begin{flalign}
\gamma_{k_j} =\frac{L_{k_j}}{2}~, \;\; \quad \quad  \Gamma_{k_j}=\Big[\omega_{n,k_j}^2(t) - \gamma^2_{k_j}\Big]^{\frac{1}{2}}\equiv\Omega_{k_j,n}(t) ~. \label{(15)}
\end{flalign}
So that Eq.~(\ref{(15)}) agrees with Eq.~(\ref{(11)}) and in both, $\Omega_{k_j,n}(t)$ is given by
Eq.~(\ref{(5)}).

By doubling the degrees of freedom, as required by the completeness of the basis $\{e^{-p_j\, t}, \, e^{+p_j t} \}$,  and introducing terms from the summation in Eq.~(\ref{(8)}) as solutions of Eq.~(\ref{(3b)}), one obtains the  same equations for the coefficients as in Eqs.~(\ref{(11)}) and (\ref{(15)}), with the same common frequency $\Omega_{k_j,n}(t)$, which thus establishes the (resonating)  link between the damped/amplified systems for each $k$-mode and a particular $n$ (cf. Section II).
We are now ready to work out explicit examples.

\section{Examples of computations and limit cycle
}

The motivation of the present Section is to show that the representation in terms of parametric damped/amplified oscillator couples leads to results which are in agreement with the ones obtained by use of ODE~\cite{Freeman1975}.
To this aim, we study two explicit examples from Ref.~\cite{Freeman1975}, Section 2.5.3. We then discuss briefly the existence of limit cycle. Since computations are straightforward we summarize them in the Appendix B.

The first example we consider is the one of the average evoked potential from the prepyriform cortex of a cat. It is fitted by means of a non-linear regression to yield the following equation (KII set)~\cite{Freeman1975}:
\begin{flalign}
& v(t) = V_{k_1}[\sin(\Gamma_{k_1}t + \varphi_{k_1})e^{-\gamma_{k_1}t} - \sin(\varphi_{k_1})e^{-\beta_{k_1} t}] +  \nonumber \\
& +V_{k_2}[\sin(\Gamma_{k_2}t + \varphi_{k_2})e^{-\gamma_{k_2}t} - \sin(\varphi_{k_2})e^{-\beta_{k_2} t}]~, \label{(16)}
\end{flalign}
where each term corresponds to a particular value of momentum $k_{j}$, $j = 1,2$, associated to each mode. $V_{k_{1,2}}$ are constants, the $\Gamma$'s and $\gamma$'s correspond to the frequency and decay rate of each component respectively, $\beta$'s are the rates of change of the initial peak and their values can be found experimentally. The interesting aspect of this particular example is that one can postulate that the $\beta$ value is in fact slightly different for the dominant and the subsidiary component (1 and 2, respectively) and each of these components is a solution for the same $k$-value, as it is the damping sinusoidal term. Details of the computation in the present formalism are reported in the Appendix B.1.

One can perform the analysis of Eq.~(\ref{(23)}) by plotting its zeros and poles in the complex plane as done in Fig.~2.
We observe that a slight change in the $\beta$ values produces changes in the shape of the fitting function (See Fig.~3).
This, together with the r\^{o}le of $\beta$ in matching the growing and decaying solutions (cf. Appendix B.1), means that the brain might use the $\beta$ parameter to discriminate between two different, though similar, behaviors or perceptions.

In the Appendix B.1 we also study the addition of two different $k$-modes with an output of the form of Eqs.~(\ref{26a}). 
We find that if we have a linear superposition of two $k$-modes of the form of Eqs.~(\ref{(16)}), where each satisfies Eq.~(\ref{(3a)}) with a different value of $\beta$, the addition of the two modes will present 4 complex poles and 2 complex zeros in $s$-space. In general, each extra mode of the same type added to the output function increases the number of complex poles and complex zeros by two when the output is represented in $s$-space.

As a second example, we consider the modeling of the synapses of axons in the primary olfactory nerve (PON)~\cite{Freeman1974} (cf. also ~\cite{Freeman1975}, Section 5.2.2). These synapses are of particular interest. The model that represents them has to account for the time delay between input and output.

We  compute the root locus plot and  can reinterpret the transfer functions in $t$-space as a linear superposition of the nonlinear $k$-modes. 
These coefficients will be given by the appropriate parameters depending on whether they are real or complex. Moreover, they can be determined by the position of the poles of the transfer function in $s$-space which are gain-dependent (Fig.~4). 

Another interesting feature is that the doubling of the degrees of freedom produces a mirror image of the poles in $s$-space with respect to the imaginary axis. The pole distribution and its variations with gain is critical when determining the chaotic dynamics of the neurological responses~\cite{Kozma2001}. An important point has to do with the ``crossing ''of the poles from the left to the right hand side in $s$-space~\cite{Freeman1975}.

\begin{figure}\label{fig:Lset}
\centering \resizebox{8cm}{!}{\includegraphics{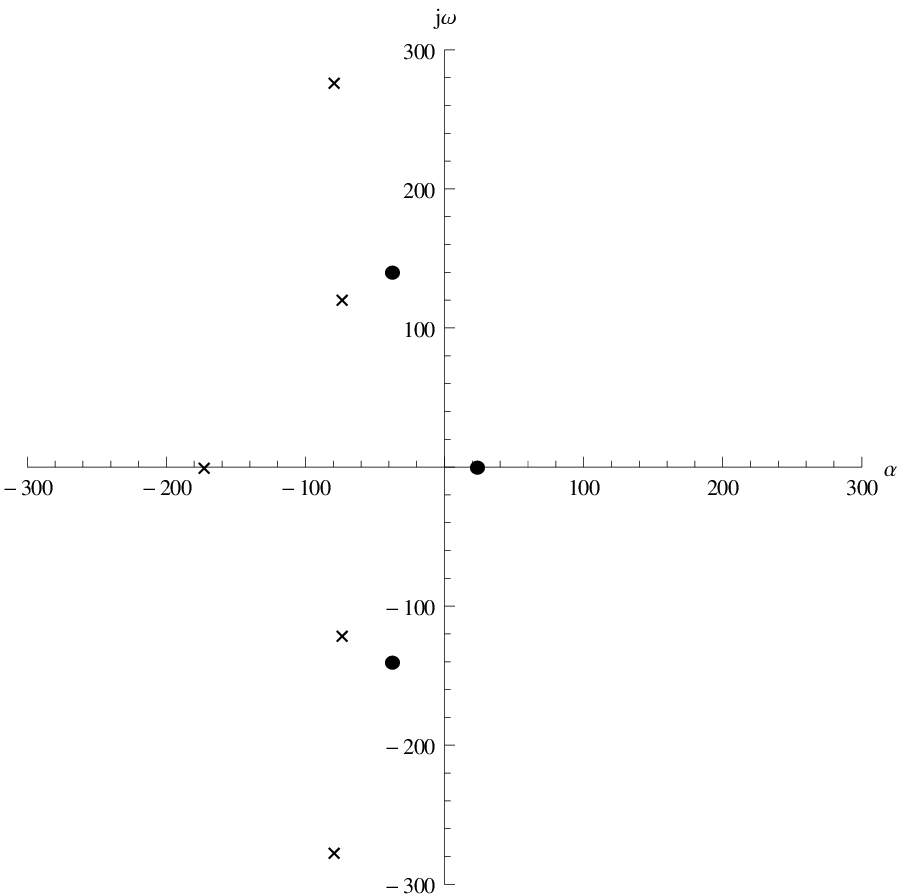}}
\caption{\small Poles ($\boldsymbol{\times}$) and zeros ($\bullet$) in $s$-space from Eq.~(\ref{(23)}).}
\end{figure}

\begin{figure}[hbtp]\label{fig:matching}
\centering \resizebox{12cm}{!}{\includegraphics{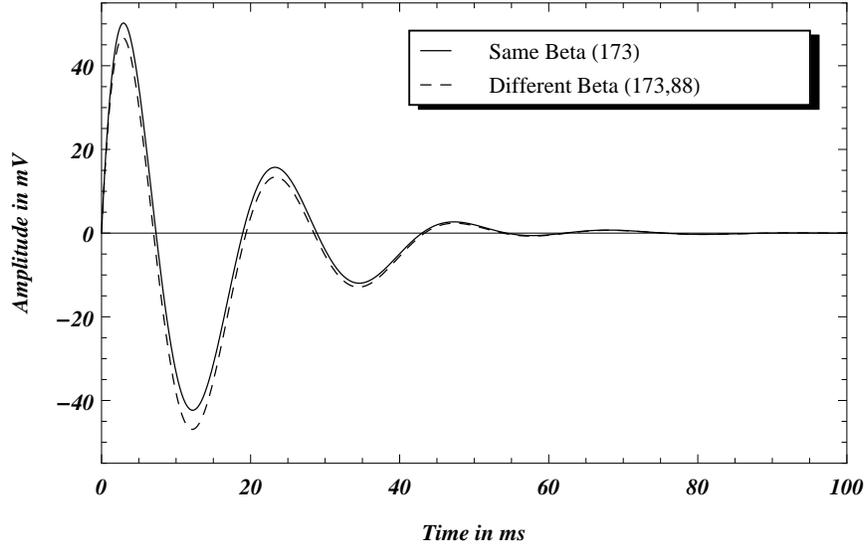}}
\caption{\small Fitting functions with same (thick/continuous) or different (dashed) $\beta$ values.
}
\end{figure}
The connections between the left and right hand side of the locus plots (i.e. damped and amplified solutions) seem to be deeply related with the chaotic dynamics that is observed in the brain~\cite{FreemanQuiran,Kozma2001}.

In the Appendix B.2 we
reproduce one of the examples in~\cite{Freeman1975} and link it to the present formalism, making thus explicit its connection with the analysis of~\cite{Freeman1975}.

The root locus plot of the transfer function $D(s)$ given by Eq.~(\ref{Ds}) is represented in Fig.~5
and one can see how  the larger poles cross the imaginary axis from the left to the right as $K_e$ decreases, which is a sign of an unstable limit cycle. The limit cycle will be stable at frequencies where the poles are purely imaginary ($\approx160 \text{ rad/sec}$).
Within our framework, the crossing of the imaginary axis means the transition from a decaying solution to an amplifying one as the gain increases.

The existence of a stable limit cycle means that the rate of oscillation of the pulses ($\Gamma_{k_j,n}$) changes with time at a fixed frequency. This agrees with our model and identifies the limit cycle frequency with $\Omega(t)$ which in fact, links the damped and amplified solutions. Therefore, the stable equilibrium state (constant pulse frequency) corresponds to fully coupled solutions ($n\rightarrow \infty$) since $\Omega(t)$ becomes time independent in this limit (see Section II and next Section). In addition, the stable limit cycle occurs for solutions with $L_{k_j}=0$.

\begin{figure}\label{fig:PST} 
\centering \resizebox{12cm}{!}{\includegraphics{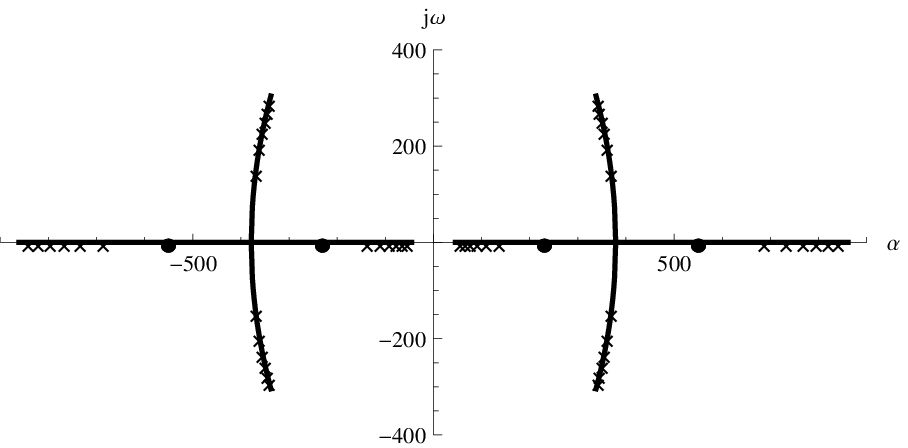}}
\caption{\small Poles ($\times$)  in root loci for $0 \le K_e \le 1$ from Eq.~(\ref{(36)}) and zeros ($\bullet$) from Eq.~(\ref{(38)}) and its ``Double''. The poles  at $s=\pm2300$ and the two zeros at $s=\pm2300$ are not shown.}
\end{figure}

\begin{figure} \label{fig:Dsplot}
\centering \resizebox{4.5cm}{!}{\includegraphics{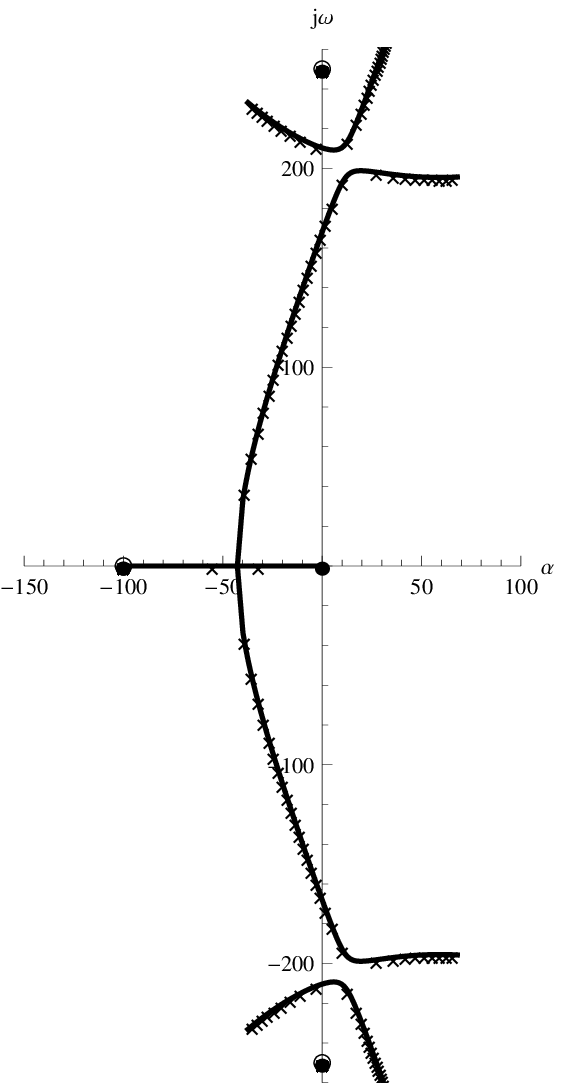}}
\caption{\small Root locus plot of Equation \eqref{Ds}.}
\end{figure}

\section{Correlation modes and their lifetime}

Our discussion has been focused on damping and amplifying terms in the fitting functions and on the root loci of zeros and poles. Now we want to discuss about lifetime and spatial range of correlation modes.

We start by recalling that the common frequency $\Omega (t)$, Eq.~(\ref{(5)}), has been assumed to be real at any time $t$ in order to avoid overdamping.  Inspection of Eq.~(\ref{(5)}) shows that such a reality condition of $\Omega (t)$ imposes an upper bound, say $T_{k_j,n}$, in the time within which a process can take place given by
\begin{flalign}
T_{k_j,n}=\frac{2n+1}{L_{k_j}}\ln\Big(\frac{2\omega_{0,k_j}}{L_{k_j}}\Big) ~. \label{(40)}
\end{flalign}
Eq.~(\ref{(40)}) imposes a bound on the admissible values of $k$: using $\omega_{0,k_j}=kc$, we have $k \ge \frac{L_{k_j}}{2c}e^{\frac{L_{k_j}}{2n+1}t} \equiv {\tilde{k}(n,\,t)}$.  This can be equivalently expressed in terms of a cut-off on the linear  range $\lambda \propto 1/{k}$ over which the $k$-mode can propagate (wave-lengths greater than ${\tilde{\lambda}} \propto 1/{\tilde{k}}(n,\,t)$ for any $n$ and $t$ are excluded).
 It it is useful to introduce the quantity  $\Lambda_{k_j,n}(t)$~\cite{Alfinito1999,AlfinitoManka2000}:
\begin{flalign}
\Lambda_{k_j,n}(t)= \frac{\ln\Big( \sinh\frac{L_{k_j}}{2n+1}T_{k_j,n}\Big)-\ln\Big(\sinh\frac{L_{k_j}}{2n+1}(T_{k_j,n}-t)\Big)+\frac{L_{k_j}}{2n+1}t}{2}~,  \label{(41)}
\end{flalign}
so that $\Lambda_{k_j,n}(0)=0 $ for any $k$ and $\Lambda_{k_j,n}(t) \rightarrow \infty $ for $t\rightarrow T_{k_j,n}$ for any $n$. Then, $\Omega(t)$ can be written as:
 \begin{flalign}
\Omega_{k_j,n}(\Lambda_{k_j,n}(t))=\Omega_{k_j,n}(0)e^{-\Lambda_{k_j,n}(t)}~.  \label{(42)}
\end{flalign}

Therefore, $\Omega_{k_j,n}(\Lambda_{k_j,n}(T_{k_j,n})) = 0$ so that $\Lambda_{k_j,n} \propto \tau_{k_j,n}\equiv$ lifetime of each $k$-mode. In this way we see that each $k$-mode ``lives" with a proper lifetime $\tau_{k_j,n}$, so that the mode is born when $\tau_{k_j,n}$ is zero and it dies for $\tau_{k_j,n}\rightarrow \infty$~\cite{AlfinitoVitiello2000}. This is illustrated in Fig.~6
for growing values of $k$ and fixed $n=1$. The figure,  where different $\Lambda_{k_j,n}$ are plotted vs $t$, shows that the lifetime increases as $k$ increases. According to this model, the modes that do not satisfy the $\Omega (t)$ reality condition are decayed modes ($\Omega_{k_j,n}=0$), which agrees with the requirement in~\cite{Freeman1975} that the damping coefficients describing the voltage output must be real and positive (cf. Eq.~(\ref{(11)})).
Since smaller $k$ implies larger $\lambda$, shorter lifetime modes condense over larger domain. Vice-versa, larger $k$ will condense on smaller domains, which, however, live longer.

The sinusoidal terms in the fitting functions are modulated by the time-dependent frequency $\Omega(t)$ so that, when the particular mode decays ($\Omega(t)=0$), the term becomes purely damping/amplifying.
On the other hand, the damping/amplifying coefficients are always expressed in terms of $L$, which we have seen is determined by intrinsic parameters of the system. In addition, the frequency $\Omega(t)$ becomes time independent when $n\rightarrow \infty$. In such a limit,  the sinusoidal terms of the fitting functions describe the system dynamical regime where energy is conserved and the modes $a_{k_j}$ fully couple to the modes $\tilde{a}_{k_j}$. As already observed,
$n$ represents the number of \emph{links} between $a$ and $\tilde{a}$ and we see that, as the number of links increases, the lifetimes of the $k$-modes are longer (Fig.~7).

Finally, the cooperative r\^{o}le of $n$ and $k$ in relation to the lifetimes is represented in Figs.~8 and 9
for ``low'' and ``high'' values of $k$ in relation to $k_0$, respectively, for growing values of $n$ (the units were chosen so that $c=1$). 

\begin{figure}[hbtp] \label{fig:fixn}
\centering \resizebox{11.5cm}{!}{\includegraphics{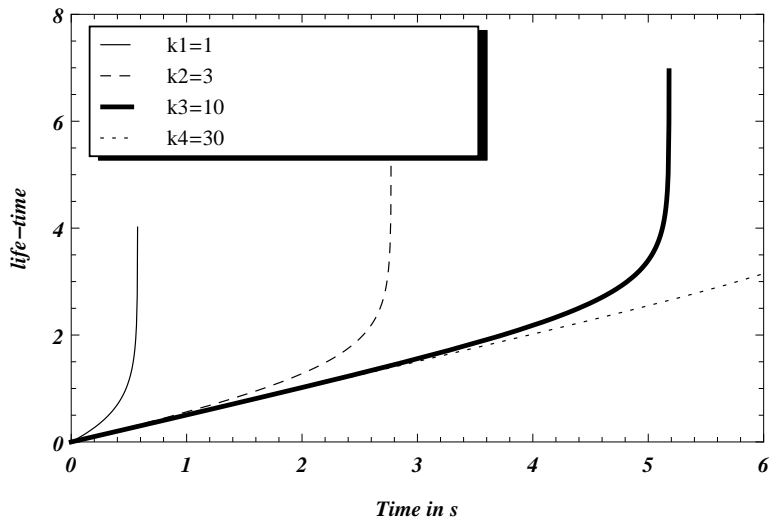}}
\caption{\small $\Lambda_{k_j,1}$ vs $t$ for different values of $k$ and fixed $n=1$.}
\end{figure}

\begin{figure}[hbtp] \label{fig:fixk}
\centering \resizebox{11.5cm}{!}{\includegraphics{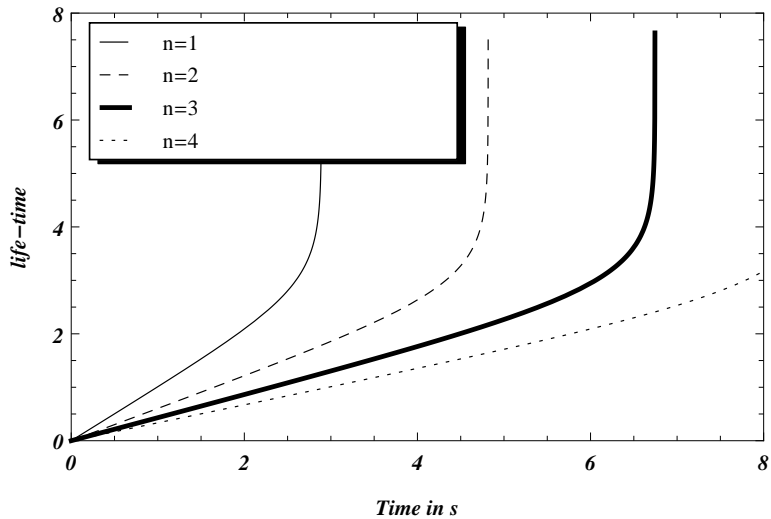}}
\caption{\small $\Lambda_{3,n}$ vs $t$ for different values of $n$ and fixed $k=3$ in units of $c=1$.}
\end{figure}

\begin{figure}[hbtp] \label{fig:3d1}
\centering \resizebox{11.5cm}{!}{\includegraphics{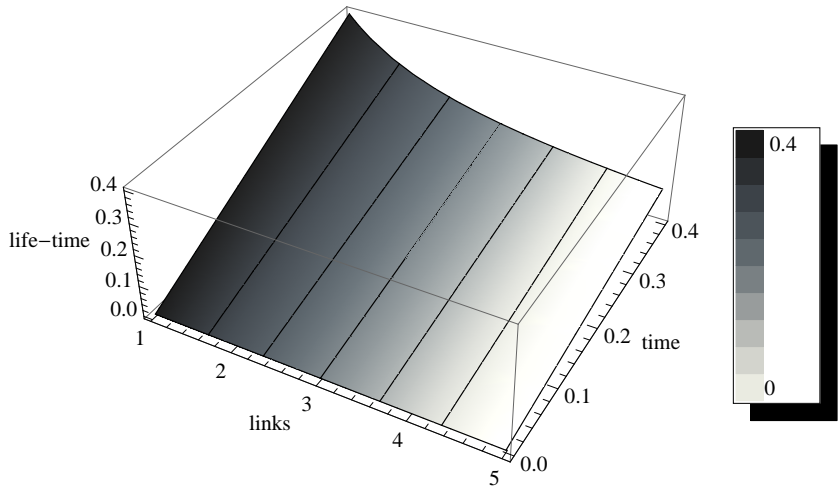}}
\caption{\small $\Lambda_{3,n}$ vs $t$ for different values of $n$ and ``low'' $k=3$.}
\end{figure}

\begin{figure}[hbtp] \label{fig:3d2}
\centering \resizebox{11.5cm}{!}{\includegraphics{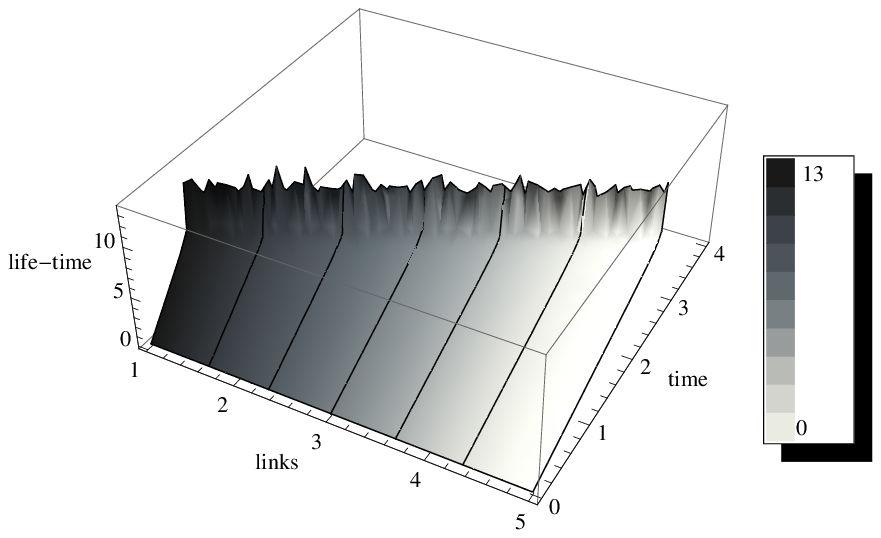}}
\caption{\small $\Lambda_{30,n}$ vs $t$ for different values of $n$ and ``high'' $k=30$.}
\end{figure}

\section{Conclusion}

The lesson we learn from the finding of the wave-pulse duality in laboratory observation of brain dynamics is that we can only progress in our discoveries if we accept both the continuum of the waves and the discreteness of the pulses and use them both, since each determines the other: pulse and wave modes coexist simultaneously in circular causality. On the other hand, it is the nonlinearity in the pulse-to-wave conversion that explains the properties of stability and instability relating to criticality and phase transitions and the variable couplings among neurons that explain reinforcement learning in populations. However, most of the time most neurons in most populations are functioning in domains of superposition, whether governed by a point attractor in a basal chaotic phase or by a limit cycle attractor in an orderly active phase.
An observer faced with understanding such a complex dynamical system is well advised to seek for a domain in which the dynamics is linear, stationary and Gaussian, as commented upon in previous Sections (and in Appendix A).
The existence of domains of superposition in the cortical dynamic state space is indeed exceedingly important. Analytic solutions of the equations for multiple negative and positive feedback loops are possible only in the phase domains where superposition holds and only when they are accessed by nonlinear phase transitions.

The small-signal linear dynamic range revealed in the observations gives us the possibility to decompose cortical impulse responses by fitting to them the sums of linear basis functions, which have been found to be solutions of ODE~\cite{Freeman1975} and have been shown in the present paper to be solutions of the couple of damped/amplified oscillator equations.
Remarkably, the damped/amplified oscillator equations are at the basis of the formulation of the dissipative quantum model of brain~\cite{MyDouble,PLR,Vitiello1995,CurrentNeuro,AlfinitoVitiello2000,FreemanVitiello2009}.  It seems thus that the very same duality particle/wave $-$ discreteness/continuum  $-$  built in in QFT upon which rests the many-body dissipative model opens a perspective in dealing with pulse-wave duality combining the complexity of nonlinear interactions with linear, stationary, ergodic solutions at the observational level. We stress that in this paper, as well as in the dissipative quantum model of brain, neurons and glia cells are considered to be classical objects. The quantum degrees of freedom are associated with the electrical dipole vibrational modes of water molecules and other molecules of biological interest.
The observational level represents the macroscopic manifestation of the underlying dynamics at the elementary constituent level~\cite{Vitiello2011}.

In this paper, the connection of the damped/amplified oscillators with the ODE formalism is made explicit by considering specific applications.
According to our discussion, the duration of a particular neuronal (``memory") process can be related to the values of the damping coefficients that describe the relevant neurological process, coefficients that can be determined from a plot of the poles of the fitting functions in $s$-space. Observations show that there are two mechanisms for determining lifetime. In the chaotic gas-like phase the lifetime is determined by the decay rate of the exponential impulse responses, which in turn is specified by the negative feedback gain. In the ordered liquid-like phase, the gain is specified by the distribution of the characteristic frequencies around the mean value of the carrier frequency~\cite{Rice,Freeman2009}.

We have that couples of damped/amplified oscillator solutions,  $\{n,\,k_j\}$ and $\{n,\,\tilde{k}_j\}$, are linked in a dissipation/absorption fashion mode.
The key of this link lies in the common value of the parameters defining the system $ \{L_{k_j}, \omega_{k_j,n}\}$. Furthermore, adding modes in pairs can be easily represented in $s$-space. With infinitely many $k$ modes, one can always pair them up.
Then, if one $k_j$ mode, for a given $j$, has a shorter lifetime than the other one in the couple, the linked system becomes unpaired (``widowed"). However, since $k_j$ and ${\tilde{k}}_j$, for each $j$, belong to continuum sets, there is also the possibility of creating new links (new couples) by selecting a new companion for the survived one. Therefore, such a possibility of recovering destroyed links could be explored.

Remarkable is also the fact that couples are among modes represented in left and right side of the root loci, which thus includes the transitions traversing the imaginary axis.
Actually, since $k_j$ (and ${\tilde{k}}_j$), for each $j$, is a variable in a continuum range,  then, we may also have dynamical couples of modes with different $i$ and $j$, for any $i$ and $j$, with continuously changeable $i$ and $j$ in their domains of variability. These would possibly describe an actively correlated neural population, as it happens in amplitude modulated neuronal sets observed in brain functioning~\cite{PLR,FreemanQuiran,FreemanVitiello2009}. In these neuronal assemblies, due to the continuous interchange of partners in the couples, nothing is "wired" to something else, we have a neuronal population made of unwired, constantly changing neuronal connections. This is truly  a dynamical assembly of correlated neurons in laminated neuropil, which is unified by an electrical interaction known to neurophysiologists as ephapsis~\cite{CapolupoPLR,Arvanitaki,Wall,Steriade,Anastassiou}, and which we conceive as forming a continuous field of exchange of information.
These assemblies are described by coherent states in the dissipative quantum model of brain (cf.~\cite{MyDouble,PLR,FreemanVitiello2009} for mathematical details). Our experimental evidence in the material domain for this ``superconducting-like'' layer is the intermittent synchronization of oscillations over the entire scalp EEG~\cite{Ruiz2010}, and it gives us
the immediate unity of rich detail in feelings and insights that constitute perception.

We finally observe that our data offer insight into the mind-body problem. The mathematics of dissipative systems requires a {\it double}, which is a mirror image defined by that portion of the universe that is engaged by the subject in action. Our data suggest that the synaptic structure of memories goes forward in time, while the remembrances in the double fed by memories recede into the past. Both structures evolve independently, yet they are ineluctably entangled at each moment of creation.

\section*{Acknowledgments}
WJF acknowledges an R01 research grant MH06686 from the National Institute of Mental Health of the United States Public Health Service. RK was supported in part by NSF CRCNS grant DMS-13-11165. AODC is grateful to the International Institute for Advanced Scientific Studies (IIASS), Vietri sul Mare (Salerno) for the grant providing partial financial support for the research presented in this paper.  AC and GV acknowledge Miur and the Istituto Nazionale di Fisica Nucleare (INFN) for partial financial support.

\appendix

\section*{APPENDICES}

\section{Empirical data, pulse density and wave density}

The empirical setting and the scenario in which the data analyzed in this paper were collected are described in this Appendix. We will comment on the linear, stationary, ergodic dynamics of the activity function observed in either of two modes, wave amplitude of dendritic current or pulse frequency, and will consider the transfer dynamics from point attractor to limit cycle attractor.

\subsection{Linear, stationary, ergodic dynamics}

Consider a population of excitatory neurons interconnected with a population of inhibitory neurons. They are distributed in a plane forming an area of cortex.  Each neuron has a branching dendritic tree providing a surface area for receiving input through synapses from $\approx 10^4$ other neurons. Each neuron has a single axon tree with multiple branches to $\approx 10^4$ other neurons.

For single neurons we define an activity function that is readily observed in either of two modes, as wave amplitude of dendritic current or a pulse frequency, depending on the mode of observation. For single cells the gold standard is a patch clamp on the surface or a microelectrode inside the cell to estimate wave amplitude of dendritic current intensity from measurements of postsynaptic potentials (EPSP and IPSP) or of the frequencies of axonal pulses (units, action potentials). There is an extensive literature (see e.g.~\cite{Granit}) showing that within a self-regulated narrow range of cortical function the conversions between wave amplitude and pulse frequency are proportional and additive, i.e., linear, for pulse to wave at synapses and wave to pulse at trigger zones.

Maintenance of that range is by a population of excitatory neurons sustaining background excitatory bias that linearizes the dynamics, giving the scale-free, temporal power spectral density, $PSD_t$, $1/f^{\epsilon}$, that characterizes the EEG spectra.

For single cells the difference is trivial. The modes have in common the real-time transmembrane voltage at the trigger zone where the axon emerges from the cell body. For populations the difference in modes is profound, because of the problems of measurement. Impulse responses are observed and measured by fitting them by nonlinear regression with sums of basis functions~\cite{Freeman1975}.
For a cortical population we define a wave density function and a pulse density function, which we sample over their time and space distributions in and around cortex. The necessity for doing so inheres in the fine structure of cortex, the neuropil. The density of cells is extreme ($10^5$ neurons$/mm^2$, area in humans of $10^{5}$ $mm^2$ and perhaps $10$ times that in glial cells), all of which contribute axons, dendrites and protoplasmic filaments between the cell bodies. With due attention paid to the architectures of cells and layers and the geometry of sources and sinks of dendritic ionic currents, we estimate the wave density from the amplitude of the local field potential (LFP) or electroencephalogram (EEG), which is the spatial sum of the contributions from all of the active neurons in the local area.

It is not possible with prevailing techniques to record  {\it simultaneously}  sufficient pulses ($\approx 10^4$) to estimate the pulse density function. Under the assumption of ergodicity and stationarity, we record long trains of pulses of single neurons, from which we estimate the relative frequency of firing by calculating the probability of firing conditional on the amplitude of the wave density~\cite{Freeman1975}. For test inputs we deliver single electric shocks to the axonal pathways to or from cortex (see Ref.~\cite{Freeman1975} for details).
The impulse input delivers all frequencies. The impulse responses (average evoked potentials (AEP) in the wave mode and post stimulus time histograms (PSTH)) in the pulse mode) display the characteristic frequencies of the cortex averaged over the time span required to collect a sample. The two modes of activity are {\it separated for observation} by band pass filtering {\it and time and space ensemble averaging}, so that the generic equation for the impulse response in time and space is the sum of the responses in the two modes of observation:

\begin{equation}\label{a}
    O(t,x,y,p,v) = O_p (t,x,y,v) + O_v (t,x,y,p),
\end{equation}
where $p$ and $v$ stand for pulse density and wave amplitudes, respectively.

By paired shock testing at differing intervals and intensities we test for domains in which additivity and proportionality hold. It turns out that~\cite{Freeman1975} test inputs that give impulse responses within the self-regulated amplitude range of the EEG reveal a small-signal linear dynamic range. Superposition enables us to decompose the cortical impulse responses by fitting to them the sums of linear basis functions, which are the solutions to linear ordinary differential equations (ODE). This finding opens a major portal into the understanding of the complicated dynamics of cortex, mainly because it facilitates analysis and modeling of the multiple feedback loops that we encounter in even the simplest forms of cortex.

\subsection{Activity is simultaneously pulse density and wave density}

With these properties the general equation to express the activity density for wave or pulse can be separated into sequential operators in time $F(t)$, space $H(x,y)$, and amplification or attenuation $G_p (v)$ at axon trigger zones or $G_v (p)$ at synapses~\cite{Freeman1975}:

\begin{eqnarray}
O_p (t,x,y,v) &=& F(t) \, G_p (v)\,  H(x,y), {\rm ~~~at~  trigger ~ zones,} \label{3}\\
 O_v (t,x,y,p) &=& F(t) \, G_v (p)\,  H(x,y), {\rm ~~~at ~ synapses.} \label{2}
\end{eqnarray}

For simplification a lumped model is developed based on a representative sample taken from a point in the population, and the spatial dimensions are subsumed by the selection and fixation of the stimulating and recording electrodes, so $H(x,y) = 1$. The intrinsically nonlinear pulse-to-wave conversion at synapses and the wave-to-pulse conversion at axonal trigger zones are replaced by coefficients,  $k_{ij}$, that express the linearized gains in the small signal range.  We find that in all instances where it is possible to measure impulse responses in the two modes, the form in both modes contains a damped cosine in which the frequency $\omega$, phase of onset $\phi$, and decay rate $\alpha$ are equal, and the amplitudes $v$ or $p$ differ only by an arbitrary constant for conversion of units of measurement which is set to one:

\begin{equation}\label{1}
    O(t) = O_v \,exp (-\alpha \,t) \, cos (\omega \,t + \phi) + O_p \, exp (-\alpha \,t) \, cos (\omega \,t + \phi)
\end{equation}

The range is extended into the nonlinear domain by piece-wise linearization and display of root loci in the complex plane of the Laplacian operator as functions of the stimulus parameters or of biological control parameters that are imposed by behavioral, physiological, surgical and pharmacological manipulations of the brain~\cite{Freeman1975}.

The wave-to-pulse nonlinear gain function was discovered to be asymmetric. Thereafter the values of the forward gains, $k_{ij}$, in the function $G_p (v)$ were determined by the slope of the tangent to the nonlinear gain curve at the operating point~\cite{FreemanQuiran}.

\subsection{ Attractor dynamics: from point attractor to limit cycle attractor}

A weak electric shock to the main input pathway to the prepyriform cortex from the olfactory bulb, which is called the lateral olfactory tract (LOT), excites a small subset of excitatory neurons. These few excite a larger number of inhibitory neurons, which in negative feedback inhibit a still larger number of excitatory neurons. They dis-excite inhibitory neurons, which dis-inhibit excitatory neurons, initiating an impulse response with the form of a damped cosine (Eq.~(\ref{1})). The impulse responses of the excitatory and inhibitory populations are in quadrature, having the same characteristic frequency but with the excitatory population leading on average  $\pi/2 ~radians$  ($90^{\circ}$). The exponential decay to the baseline is modeled by postulating a point attractor under perturbation by noise. Convergence to the pre-stimulus baseline is a necessary condition for observation of both modes. The impulse response must terminate before the next impulse is given. The decay rate,  $\alpha$, must be negative, and the characteristic frequency, $\omega$, must be given by the conjugate pair of poles closest to the imaginary axis of the complex plane.

The frequency and decay rate of the complex conjugate pair of poles is primarily determined by the negative feedback gain, $k_n$, between the excitatory population, $KI_e$, and the inhibitory population, $KI_i$. It is strongly modulated by the positive feedback gains, $k_{ee}$ and $k_{ii}$, within respectively the $KI_e$ and $KI_i$ populations.

Modeling the dynamics of the three loops can be postponed by noting the fact that the neurons in the populations get no information about what are the sources of their inputs. They are embedded in neuropil and integrate whatever synaptic input they receive from all sources. This fact can be modeled by postulating that when a neuron fires a pulse, the axon distributes the pulse to $\approx ~10^4$ other neurons, and some of those to $\approx ~10^8$ others, and so on, reverberating through the neuropil as it decays. The neural activity that feeds back to the originating neuron (or neurons) embedded in the neuropil can be modeled as a one-dimensional diffusion process. using the linear transcendental Laplacian operator, $K_2 \,exp[-(s\,T_2)^{0.5}]$, where $K_2$ is the lumped gain and $T_2$  is the lumped time delay of the distributed feedback.

The advantage of this simplified model was that it revealed a fundamental property of the interactive populations of excitatory and inhibitory neurons, modeled as the $KII_{ei}$ set.
When the impulse input intensity was fixed in resting subjects, repeated samples of $O(t)$ revealed spontaneous variation of the damped cosine. It turned out that $K_2$ and $T_2$ were linearly correlated. Variation in lumped gain $K_2$ gave root loci for $\omega$ and $\alpha$ that crossed into the right half of the complex plane, reversing the sign of $\alpha$ from attenuating to amplifying, and predicting transfer of cortical control from a point attractor to a limit cycle attractor. In physiological terms the closer cortical axons were brought by noise to their thresholds, the more likely they were to fire. This well-known biophysical nonlinearity at the microscopic level appeared at the macroscopic level as the root locus in Mode 2~\cite{Freeman1975}, thereby revealing a mechanism capable of explaining how a surge in neural activity could provide the transition energy required to initiate sustained oscillation by a phase transition resembling a Hopf bifurcation.

\section{Explicit examples of computations}

We present explicit computations in terms of Eqs.~(\ref{(3a)}) and (\ref{(3b)}) for two examples from Ref.~\cite{Freeman1975}, Section 2.5.3: the average evoked potential from KII set and poststimulus time histograms of single glomerular neurons in olfactory bulb. We also summarize the discussion on the existence of the limit cycle.
Preliminary results have been presented in~\cite{Olivares}.

\subsection{Average evoked potential from KII set}

We refer to Eq.~(\ref{(16)}) and will drop the $k_{j}$ suffix for convenience, keeping only the label of each momentum in order to repeat the analysis as in \cite{Freeman1975}, assuming that the $\beta$'s are common for each mode (i.e. $\beta_1=\beta_2$).

The prepyriform response to a $\delta(t)$ input can be modeled by 2 parallel filters where each is cascaded into a lead-lag filter~\cite{Freeman1975}. This means that the response function $c(t)=c_1(t) + c_2(t)$ has one real zero, one real pole and two complex conjugates poles per component, which can be written in the $s$ space as
\begin{flalign} \label{(18)}
 C(s)=C_1(s)+C_2(s)~,  \qquad
 \text{with}\;\;C_j=\frac{K_j (s + z_j)}{(s+ B_j)(s+B^*_j)(s+\beta)}, \;\quad j=1,2 ~,
\end{flalign}
where the $K$'s are constants, the $\gamma$'s can be found by graphical analysis and determine the phase of the output.
By partial fraction expansion, Eq.~(\ref{(18)}),
when separated in modulus and phase, yields
\begin{flalign}
& C_1(s)= \frac{K_1}{\Gamma_1}\Big[\frac{\Gamma_1^2 + (z_1 - \gamma_1)^2}{\Gamma_1^2 + (\beta-\gamma_1)^2}\Big]^{\frac{1}{2}}\Big( \frac{e^{i\theta_1}}{2(s+B_1)} + \frac {e^{-i\theta_1}}{2(s+B_1^*)}+ \nonumber\\
& + \frac{\Gamma_1(z_1 - \beta)}{[\Gamma_1^2 + (z_1 - \gamma_1)^2]^{\frac{1}{2}}[\Gamma_1^2 + (\beta-\gamma_1)^2]^{\frac{1}{2}}(s+\beta)}\Big)~,  \label{(20)}
\end{flalign}
with $\theta_1=tan^{-1}(Im(C_1(s))/{Re(C_1(s))})$. The same result applies to $C_2(s)$.

Since we are dealing with a $\delta(t)$ input and $c(t)$ output, we have the input $V(s)=C(s)$ so that use of the Laplace transform 
gives
\begin{flalign}
& C_1(s)=V_1(s) = V_1[\frac{A_1}{s + B_1}+ \frac{A_1^*}{s+B_1^*}- \frac{D_1}{s + \beta}] \label{(21)} ~,
\end{flalign}
and similarly for $C_2(s)$.
This allows us to compare the experimental data described by Eq.~(\ref{(16)}) with the modeled output (Eq.~(\ref{(18)})) and find the values of $K_{1,2}$. Comparison of Eq.~(\ref{(20)}) and Eq.~(\ref{(21)}) shows that:
\begin{flalign}
&K_1= V_1\Gamma_1 \Big[\frac{\Gamma_1^2 + (\beta - \gamma_1)^2}{\Gamma_1^2 + (z_1-\gamma_1)^2}\Big]^{\frac{1}{2}} \label{(22)}~, \\
& \sin({\varphi_1})=\frac{\Gamma_1(\beta-z_1)}{[\Gamma_1^2 + (z_1 - \gamma_1)^2]^{\frac{1}{2}}[\Gamma_1^2 + (\beta-\gamma_1)^2]^{\frac{1}{2}}} ~, \qquad
\theta_1=-(\varphi_1 - \frac{\pi}{2})~,
\end{flalign}
with an equivalent expression for the  $K_2$-term.
 Then, the full expression of $C(s)$ can be written as
\begin{eqnarray}
C(s) &=& \frac{(K_1 + K_2)s^3 + P_1s^2 + P_2s + P_3}{(s+B_1)(s+B_1^*)(s+B_2)(s+B_2^*)(s + \beta)} \label{(23)} ~,\\
\text{with}\; P_1 &=& K_1(z_1 +2\gamma_2) + K_2 (z_2 +2\gamma_1)~, \\
 P_2 &=& K_1(2\gamma_2 z_1 + \gamma_2^2 + \Gamma_2^2) + K_2(2\gamma_1 z_2 + \gamma_1^2 + \Gamma_1^2)~,\\
 P_3 &=&K_1 z_1(\gamma_2^2 + \Gamma_2^2) + K_2 z_2(\gamma_1^2 + \Gamma_1^2)~.
\end{eqnarray}

Zeros and poles of Eq.~(\ref{(23)}) in the complex plane  are plotted in Fig.~2.
We can regard each damping/sinusoidal-damping term in Eq.~(\ref{(16)}) as the observable excitations of a particular $k$-mode, being the modeled output a linear superposition of different $k$-modes. Moreover, we will postulate the double of this system as an amplified version of Eq.~(\ref{(16)}) :
\begin{flalign}
& v(t) = V_{k_1}[\sin(\Gamma_{k_1}t + \varphi_{k_1})e^{\gamma_{k_1}t} - \sin(\varphi_{k_1})e^{-\beta_{k_1} t}] +  \nonumber\\
& +V_{k_2}[\sin(\Gamma_{k_2}t + \varphi_{k_2})e^{\gamma_{k_2}t} - \sin(\varphi_{k_2})e^{-\beta_{k_2} t}] ~, \label{(24)}
\end{flalign}
where the $\beta_{k_j}$ are the same for the amplified and the damped system.

Each coefficient will satisfy the restrictions given by Eqs.~(\ref{(11)}) - (\ref{(15)}). Moreover, by requiring that each $\beta$ value is a solution of the same $k$-mode as the sinusoidal damping/amplifying term, when substituting the $\beta_{k_{j}}$ term into Eqs.~(\ref{(3a)}) and (\ref{(3b)}) we obtain:
\begin{flalign} \label{(25a)}
 \beta_{k_j,n}= \gamma_{k_j,n} \pm i \Gamma_{k_j,n}~, \qquad
 \beta_{\tilde{k}_j,n}= - \gamma_{\tilde{k}_j,n} \pm i \Gamma_{\tilde{k}_j,n}~. 
\end{flalign}
Thus, in order for  $\beta_{k_j}$ to be common  to (i.e. the link between) the damped/amplified pair, it has to be $\gamma_{k_1,n}=- \gamma_{\tilde{k}_1,n}$ and $\Gamma_{k_1,n}= \Gamma_{\tilde{k}_1,n}$,  which agrees with the previous discussion. Therefore, $\beta$ (which is always a damping coefficient) is indeed a link between the damped/amplified coupled mode.

If one substitutes each value of $\beta$ for a single $k_j$ mode in Eq.~(\ref{(16)}), (dropping the $k$ and $n$ subscripts again for simplicity) the solutions reduce to a single damped/amplified sinusoidal term with a phase factor:

 \begin{flalign} \label{26a}
  V (t) = \sin (\Gamma t) e^{-\gamma t} e^{\pm i \varphi}   ~, \qquad
 \tilde{V}(t) = \sin (\Gamma t)e^{\gamma t} e^{\pm i \varphi}~. 
\end{flalign}

We can now repeat the analysis of Ref.~\cite{Freeman1975} for the addition of two different $k$-modes but with an output of the form of Eqs.~(\ref{26a}). 
Note that, according to the discussion in \cite{Freeman1975}, the absence of a decaying $\beta$ coefficient implies the cancellation of the real zero and the real pole in the transmission function of the output in $s$-space (i.e. $z=\beta$ for each $k$-mode) so that the model simply reduces to two parallel band-pass filters which are not cascaded into a lead-lag filter, represented by 2 complex conjugates poles in $s$-space.
The modeled output $c(t)=c_{k_1}(t) + c_{k_2}(t)$ leads finally to
 \begin{flalign}
 &C(s)= C_1(s) + C_2(s) = \nonumber \\
 &= \frac{(K_1+K_2)s^2+(K_12\gamma_2 + K_22\gamma_1)s+K_1(\gamma_2^2 + \Gamma_2^2)+K_2(\gamma_1^2 + \Gamma_1^2)}{(s + \gamma_1 + i \Gamma_1)(s + \gamma_1 - i \Gamma_1)(s + \gamma_2 + i \Gamma_2)(s + \gamma_2 - i \Gamma_2)} \label{(33)}
 \end{flalign}
in $s$-space.
The numerator of Eq.~(\ref{(33)}) is a quadratic equation with two complex conjugate roots. In fact, one can notice the resemblance with Eqs.~(\ref{(3a)}) and (\ref{(3b)})  by using the definitions of $L$ and $\omega$:
 \begin{flalign}
 & (K_1+K_2)s^2+(K_1L_2 + K_2L_1)s+(K_1\omega_2^2+K_2\omega_1^2) = 0 ~.\label{(34)}
 \end{flalign}

Therefore, we conclude that a linear superposition of two $k$-modes of the form of Eqs.~(\ref{(16)}), where each satisfies Eq.~(\ref{(3a)}) with a different value of $\beta$, leads to 4 complex poles and 2 complex zeros in $s$-space. Then, each extra mode of the same type added to the output function will increase the number of complex poles and complex zeros by two when the output is represented in $s$-space. In addition, when the amplified terms are also considered (solutions of Eq.~(\ref{(3b)})), its poles and zeros are a reflection in the imaginary axis of the ones presented by Eq.~(\ref{(33)}).

\subsection{Poststimulus time histograms of single glomerular neurons in olfactory bulb}

The second example we consider is the modeling of the synapses of axons in the primary olfactory nerve (PON)~\cite{Freeman1974} (cf. also ~\cite{Freeman1975}, Section 5.2.2).

Starting from a  model based on a characteristic set of differential equations,  the transfer function for the feedback subset of the closed loop representing these synapses is~\cite{Freeman1974}
 \begin{flalign}
 &P_c(s)=\frac{K_cK_e \,8.5\times10^{16}(2300-s)^2}{(s+\gamma +i \Gamma)(s + \gamma - i\Gamma)(s + p_1)(s + p_2)(s + p_3)(s + p_4)}~, \label{(36)}
 \end{flalign}
where all the coefficients are real and positive, $K_c$ is the closed loop gain, $K_e$ is the feedback gain and the pole coefficients are experimentally determined.
The inverse transform of Eq.~(\ref{(36)}) is
 \begin{flalign}
 &p_c(t ) = K_g [P_1e^{-p_1 t}+ P_2 e^{-p_2 t}
+ P_3 e^{-p_3 t} + P_4 e^{-p_4 t}
+ P_5 \sin(\Gamma t + \varphi)e^{-\gamma t}] ~,\label{(37)}
 \end{flalign}
where the $P$'s are parameters to be determined and $K_g$ is the overall gain. The validity of Eq.~(\ref{(36)}) is confirmed by the similarity of Eq.~(\ref{(37)}) with the recorded poststimulus time (PST) histograms.
The overall transfer function is 
 \begin{flalign}
 &P_q(s)=\frac{K_gK_e \, 8.5\times10^{16}(s+z_1)(s+z_2)(2300-s)^2}{(s+\gamma +i \Gamma)(s + \gamma - i\Gamma)(s + p_1)(s + p_2)(s + p_3)(s + p_4)(s+p_5)(s+p_6)} ~,\label{(38)}
 \end{flalign}
where
$z_1,z_2=\frac{-(p_5+p_6)\pm d}{2}$, respectively, with $d=\big[(p_5+p_6)^2 - 4p_5p_6(1+K')\big]^{\frac{1}{2}}$.
$K'$ is the amplitude of the surge relative to the pulse.
Note that we can calculate the zeros of Eq.~(\ref{(38)}) from $z_1,z_2$ and the poles will change with $K_e$ so that we can compute the root locus plot. Again, the transfer function is
 \begin{flalign}
 &p_q(t ) = K_g [P_1e^{-p_1 t}+ P_2 e^{-p_2 t}
+ P_3 e^{-p_3 t} + P_4 e^{-p_4 t}
+ P_5 \sin(\Gamma t + \varphi)e^{-\gamma t} + \nonumber \\
& \;\;\;\;\;\;\;\;\;\;+ P_6e^{-p_5} + P_7e^{-p_6} ]~, \label{(39)}
 \end{flalign}
which is very similar to the recorded PST histograms.

The transfer functions in $t$-space are then interpreted as a linear superposition of the nonlinear $k$-modes.
The position of the poles of the transfer function in $s$-space which are gain-dependent determine the coefficients in the superposition (Fig.~4). 

\subsection{Limit cycle discussion}

The transfer function $D(s)$ for a KIII set output from a KII set with negative feedback from a KI set is given by:
\begin{equation} \label{Ds}
D(s)= \frac{C(s)}{1+K_eC(s)P(s)}~,
\end{equation}
where $C(s)={(6.25 \times 10^{6})}/{(s+i250)(s-i250)(s+100)}$ and $P(s)=100/{s(s+100)}$ are, respectively, the KII and KI set transfer functions. 
As $K_e$ changes, the poles of Eq.~(\ref{Ds}) change accordingly:
\begin{equation}
D(s)= \frac{6.25 \times 10^6 s (s+100)}{(s+p_1)(s+p_2)(s+p_3)(s+p_4)(s+p_5)}~,
\end{equation}
where the $p_n$ poles can be real or complex. The $D(s)$ root locus is plotted in Fig.~5. 
The larger poles cross the imaginary axis from the left to the right as $K_e$ decreases. This signal an unstable limit cycle. At frequencies corresponding to purely imaginary poles ($\approx160 \text{ rad/sec}$) the limit cycle is stable.

\end{document}